\documentclass[Times,4pt,preprint,twocolumn,superscriptaddress,2pt]{revtex4-1}
\usepackage{amsmath,natbib,graphicx,subfigure,amssymb,graphics,amsmath,mathrsfs,CJK,color}
\usepackage{multirow,fancyhdr,color,bm,tabularx,psfrag,geometry,dcolumn,datetime}
\usepackage[colorlinks=true,bookmarksopen,bookmarksnumbered,citecolor=blue, linkcolor=blue, urlcolor=blue]{hyperref}
\def\footnoterule{\kern -1mm \hrule width 6.6cm \kern 2.2mm}%
\usepackage{tikz,xcolor,hyperref}
\definecolor{lime}{HTML}{A6CE39}
\DeclareRobustCommand{\orcidicon}{%
    \begin{tikzpicture}
    \draw[lime, fill=lime] (0,0)
    circle [radius=0.16] node[white]
   {{\fontfamily{qag}\selectfont \tiny ID}};\draw[white, fill=white] (-0.0625,0.095)
    circle [radius=0.007];
    \end{tikzpicture}
    \hspace{-2mm}}
\foreach \x in {A, ..., Z}
{\expandafter\xdef\csname orcid\x\endcsname{\noexpand\href{https://orcid.org/\csname orcidauthor\x\endcsname}{\noexpand\orcidicon}}}

\begin{document}

\title{\LARGE Predicting quantum evolutions of excitation energy transfer in a light-harvesting complex using multi-optimized recurrent neural networks}
\author{Shun-Cai Zhao\orcidA{}}
\email[Corresponding author: ]{zsczhao@126.com.}
\affiliation{Department of Physics, Faculty of Science, Kunming University of Science and Technology, Kunming, 650500, PR China}
\affiliation{Center for Quantum Materials and Computational Condensed Matter Physics, Faculty of Science, Kunming University of Science and Technology, Kunming, 650500, PR China}

\author{Yi-Meng Huang}
\email[Co-first author: ]{zscgroup@kmust.edu.cn.}
\affiliation{Department of Physics, Faculty of Science, Kunming University of Science and Technology, Kunming, 650500, PR China}
\affiliation{Center for Quantum Materials and Computational Condensed Matter Physics, Faculty of Science, Kunming University of Science and Technology, Kunming, 650500, PR China}

\author{Zi-Ran Zhao}
\affiliation{Department of Physics, Faculty of Science, Kunming University of Science and Technology, Kunming, 650500, PR China}
\affiliation{Center for Quantum Materials and Computational Condensed Matter Physics, Faculty of Science, Kunming University of Science and Technology, Kunming, 650500, PR China}

\begin{abstract}
Constructing models to discover physics underlying magnanimous data is a traditional strategy in data mining which has been proved to be powerful and successful. In this work, a multi-optimized recurrent neural network (MRNN) is utilized to predict the dynamics of photosynthetic excitation energy transfer (EET) in a light-harvesting complex. The original data set produced by the master equation were trained to forecast the EET evolution. An agreement between our prediction and the theoretical deduction with an accuracy of over 99.26\% is found, showing the validity of the proposed MRNN. A time-segment polynomial fitting multiplied by a unit step function results in a loss rate of the order of $10^{-5}$, showing a striking consistence with analytical formulations for the photosynthetic EET. The work sets up a precedent for accurate EET prediction from large data set by establishing analytical descriptions for physics hidden behind, through minimizing the processing cost during the evolution of week-coupling EET.
\begin{description}
\item[Keywords]{Quantum evolution, excitation energy transfer, multi-optimized recurrent neural networks, quantum dynamics}
\end{description}
\end{abstract}

\maketitle
\section{Introduction}

Many physical laws were discovered based on data, such as Kepler's three laws via large data of celestial body motion observed by Tycho. Understanding the temporal evolution of photosynthetic excitation energy transfer (EET) in a light-harvesting complex is an important topic of broad interest due to its nearly 100\% photosynthetic conversion efficiency, providing an ideal option for mitigating energy crisis\cite{AlharbiTheoretical,LiZhao2021,li2021chargetransport}. Exact numerical simulations of the dynamics of EET in a light-harvesting complex, on the other hand, requires enormous computational resources\cite{1ccc,huo2015electronic}, which tends to grow exponentially with the number of simulated time steps and system size. Even though many techniques, such as the hierarchy of equations of motion (HEOM) technique\cite{strmpfer_2012_open,tanimura2020numerically}, multi-configurational time-dependent Hartree (MCTDH)\cite{stockburger_2002_exact}, stochastic Liouville-von Neumann equation\cite{meyer_1990_the}, quasi-adiabatic propagator path-integral (QUAPI)\cite{makri1998quantum}, and path-integral Monte Carlo\cite{kast2013persistence} are available, they are inappropriate for examining long-time quantum dynamical evolution.

The current state of a quantum system is mostly determined by and rooted in its early stages of evolution, which enables us to learn long-time evolution of EET in a light-harvesting complex from short-time dynamics without the costly and direct long-time simulations. Once a memory kernel is acquired, the Nakajima-Zwanzig generalized quantum master equation (GQME)\cite{1958On} provides a broad and formally reliable prescription for achieving this goal\cite{2013Efficient}. Nonetheless, solving the GQME and directly computing the memory kernel for an arbitrary system is still challenging. The transfer tensor method (TTM) can solve the GQME somehow, but it requires an external numerical methodology to provide a set of dynamical mappings\cite{PhysRevLett.112.110401,2016Accurate,2020Non}. The interplay between machine learning and quantum physics, on the other hand, altered the current situation\cite{naicker2022machine,2021Machine,lin2022automatic} by providing new concepts for modeling the evolution of EET in a light-harvesting complex\cite{0Predicting}, i.e. intuitively and directly learning from data set without explicit theoretical construction. Artificial neural networks (ANNs)\cite{2018Modelling} have shown that complex functional dependencies in time series can be learned directly from data\cite{2020Learning,2020Convolutional}. This evades the great efforts to make theoretical analyses, which sometimes become unjustified (e.g., the weak coupling limit) or difficult to derive if just based on phenomenological observations. ANNs have been demonstrated to be capable of solving the master equations governing the dynamics of long-time dissipative open quantum systems\cite{PhysRevLett.122.250502}.

The recurrent neural network (RNN) in particular has an exceptional capacity to interpret an intricate temporal behavior. It can preserve historical information for future purpose of prediction by building a feedback loop that receives both the current stage's input and the previous step's output\cite{LEM0Machine}. However, concern of gradient vanishing or exploding behavior due to multiple iterating operations, on the other hand, limits the use of RNNs in long-time scale applications\cite{2016Army}. To address this flaw, this work employs multi-optimized recurrent neural networks (MRNNS) rather than long short-term memory recurrent neural networks (LSTM-NN)\cite{hochreiter1997long,lin2021simulation}, to model the long-term dependencies of the time-series data set, by storing the key information, and predicting the future data that is not currently available. MRNNs are also used as propagators of the time-dependent master equations to regulate the light-harvesting complex over a range of time scales.

The rest of the paper is organized as follows. The quantum processes and the common master equations for photosystem II reaction center (PSII-RC) are described in Sec.2.A, and a sample RNN architecture with optimized hyperparameters is introduced in Sec.2.B. Results are discussed in Sec.3, where we validate the learning model in the interval [0.~80]fs (Sec.3.A) and predict the EET evolution process in the range [80,~500]fs (Sec.3.B), using the polynomial fitting and the analytical expression(Sec.3.C). Finally, conclusions and outlook of the this work are summarized in Sec.4.

\section{Theoretical model and multi-optimized recurrent neural networks}
\subsection{Theoretical model for Photosystem II reaction center (PSII-RC)}

\begin{figure}
\center
\includegraphics[width=0.70\columnwidth]{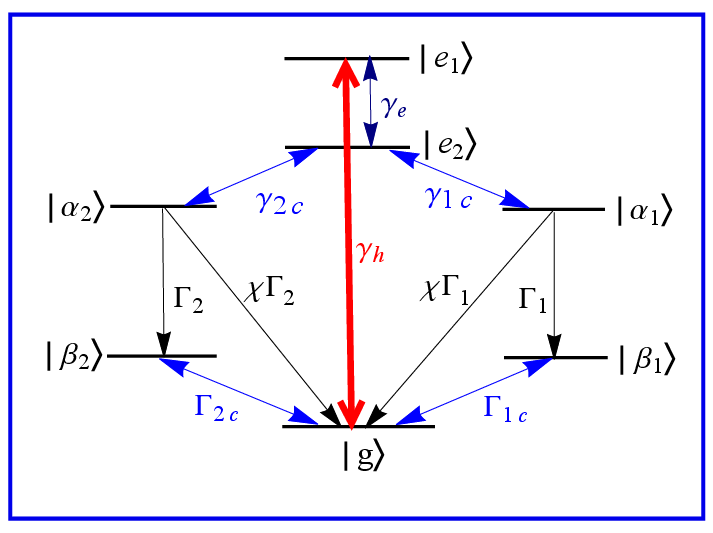 }
\caption{{\scriptsize(Color online) Energy-level framework for the photosynthetic RC with two load-transitions \(|\alpha_{i}\rangle\)\(\rightarrow\)\(|\beta_{i}\rangle_{i=1,2}\). The electronic transitions from ground state \(|g\rangle\) to two coupled dipoles \(|e_{1}\rangle\) and \(|e_{2}\rangle\) are induced by the high temperature photon bath, while the low temperature phonon bath drives charge transfer from \(|e_{2}\rangle\) to \(|\alpha_{i}\rangle_{i=1,2}\), and \(|\beta_{i}\rangle_{i=1,2}\) to \(|g\rangle\) with a termination of electronic circulation. }}
\label{fig1}
\end{figure}

A typical photosystem II reaction center (PSII-RC) often seen in purple bacteria and oxygen-evolving organisms (cyanobacteria, algae, and higher plants) comprises six pigment molecules located at the core of the complex with two symmetric branches of protein matrix\cite{NovoderezhkinMixing}. Four chlorophylls of them (special pair PD\(_{1}\), PD\(_{2}\) and accessory ChlD\(_{1}\), ChlD\(_{2}\)) and two pheophytins (PheD\(_{1}\), PheD\(_{2}\)), are parallel distributed in these two branches of protein matrix\cite{ElisabetTwo}. The pair of chlorophylls, PD\(_{1}\)and PD\(_{2}\), located at the center of the PSII RC act as the primary electron donors, forming two excited states denoted as \(|e_{1,2}\rangle\) in Fig.\ref{fig1}. Two pheophytin pigments, PheD\(_{1}\) and PheD\(_{2}\) couple to the rest of the molecules and act as the electron acceptors\(|\alpha_{1}\rangle\) and \(|\alpha_{2}\rangle\), respectively\cite{Novoderezhkin2015Multiple}, as shown in Fig.\ref{fig1}. This is an energy-level framework abstracted from the PSII-RC\cite{LiZhao2021}, where \(|\beta_{1}\rangle\)(\(|\beta_{2}\rangle\)) is a positively charged state after an electron is released and \(|g\rangle\) is a ground state.

After absorbing a solar photon, an electron is excited from \(|g\rangle\) to \(|e_{2}\rangle\) with a transition rate \(\gamma_{h}\), where the excited electron may transit to \(|e_{1}\rangle\) state at a rate \(\gamma_{e}\). Then the excited electron is transferred to the acceptors by emitting a phonon via two pathways: \(|e_{1}\rangle\)(\(|e_{2}\rangle\))\(\rightarrow\) \(|\alpha_{1}\rangle\) or \(|e_{1}\rangle\)(\(|e_{2}\rangle\))\(\rightarrow\) \(|\alpha_{2}\rangle\) at emission rates \(\gamma_{1c}\) and \(\gamma_{2c}\), respectively, with \(|\alpha_{1}\rangle\) and \(|\alpha_{2}\rangle\) being the ion-pair states in these two pathways. Such a process is accompanied by a spatial separation of positive and negative charges induced by the release of the excited electrons to the plastoquinone molecule, leaving a hole in the dimer. Then the electrons are released from the two acceptors PheD\(_{1}\) and PheD\(_{2}\) denoted by \(|\alpha_{1}\rangle\) \(\rightarrow\) \(|\beta_{1}\rangle\) (Path 1) and \(|\alpha_{2}\rangle\) \(\rightarrow\) \(|\beta_{2}\rangle\) (Path 2) with respective rate \(|\Gamma_{i}\rangle_{(i=1,2)}\), providing energies for possible work. Finally, the electron returns to the primary electron donor via \(|\beta_{1}\rangle\)(\(|\beta_{2}\rangle\)) \(\rightarrow\) \(|g\rangle\). As an alternative pathway to the two-step processes \(|\alpha_{1}\rangle\)(\(|\alpha_{2}\rangle\)) \(\rightarrow\)\(|\beta_{1}\rangle\)(\(|\beta_{2}\rangle\)) \(\rightarrow\) \(|g\rangle\), the acceptor-to-donor charge recombination can also be made by directly bringing the system back to the ground state \(|g\rangle\) without producing any current with rates \(\chi \Gamma_{i=1,2}\), where \(\chi\) is a dimensionless fraction\cite{LiZhao2021} describing the radiative recombination rate of the two pathways.

In order to learn the long-time behaviors of EET in the above-mentioned PSII-RC, in stead of the costly and direct long-time simulations, we employ MRNNs to predict its evolution. In this work,the learning model starts with the data collected from the initial stage of the PSII-RC, which is generated by a weak contact of the system with its environment.

The unitary evolution of the electron transfer, as usual, can be described by an electronic Hamiltonian

\begin{eqnarray}
&\hat{H}_{e}=&E_{g}|g\rangle\langle g|+\sum^{2}_{i=1}(E_{\alpha_{i}}|\alpha_{i}\rangle\langle \alpha_{i}|+E_{\beta_{i}}|\beta_{i}\rangle\langle \beta_{i}| \nonumber\\
             &&+E_{e_{i}}|e_{i}\rangle\langle e_{i}|)+ (|e_{1}\rangle\langle e_{2}|+|e_{2}\rangle\langle e_{1}|),                          \label{eq0}
\end{eqnarray}
via a Lindblad-type master equation
\begin{eqnarray}
&\frac{d\hat{\rho}}{dt}=&-i[\hat{H}_{e},\hat{\rho}]+\mathscr{L}_{H}\hat{\rho}+\mathscr{L}_{1c}\hat{\rho}+\mathscr{L}_{2c}\hat{\rho}+\mathscr{L}_{3c}\hat{\rho} \nonumber\\
                       &&+\mathscr{L}_{\Gamma}\hat{\rho}+\mathscr{L}_{\lambda\Gamma}\hat{\rho},                                             \label{eqa}
\end{eqnarray}

\noindent where the strength of the interaction with the environment is comparable with the internal interactions inside the system. And the last term in Eq.~(\(\ref{eq0}\)) shows the the dipole-dipole coupling between \(|e_{1,2}\rangle\) is set as one unit. Then the Lindblad-type superoperators in Eq.~(\ref{eqa}) are listed as

{\footnotesize
\begin{widetext}
\begin{eqnarray}
&\mathscr{L}_{H}\hat{\rho}=&\frac{\gamma_{h}}{2}[(n_{h}+1)(2\hat{\sigma}_{g1}\hat{\rho}\hat{\sigma}_{g1}^{\dag}-\hat{\sigma}_{g1}^{\dag}\hat{\sigma}_{g1}
                             \hat{\rho}-\hat{\rho}\hat{\sigma}_{g1}^{\dag}\hat{\sigma}_{g1})\nonumber\\
                           &&+n_{h}(2\hat{\sigma}_{g1}^{\dag}\hat{\rho}\hat{\sigma}_{g1}-\hat{\sigma}_{g1}\hat{\sigma}_{g1}^{\dag}\hat{\rho}-\hat{\rho}\hat{\sigma}_{g1}\hat{\sigma}_{g1}^{\dag})],\label{eqb}\\
&\mathscr{L}_{1c}\hat{\rho}=&\sum^{2}_{i,j=1}\frac{\gamma_{ijc}}{2}[(n_{1c}+1)(\hat{\sigma}_{j2}\hat{\rho}\hat{\sigma}_{i2}^{\dag}
                             +\hat{\sigma}_{i2}\hat{\rho}\hat{\sigma}_{j2}^{\dag}-\hat{\sigma}_{j2}^{\dag}\hat{\sigma}_{i2}\hat{\rho}-\hat{\rho}\hat{\sigma}_{i2}^{\dag}\hat{\sigma}_{j2})\nonumber\\
                            &&+n_{1c}(\hat{\sigma}_{i2}^{\dag}\hat{\rho}\hat{\sigma}_{j2}+\hat{\sigma}_{j2}^{\dag}\hat{\rho}\hat{\sigma}_{i2}-\hat{\sigma}_{i2}\hat{\sigma}_{j2}^{\dag}\hat{\rho}-\hat{\rho}\hat{\sigma}_{j2}\hat{\sigma}_{i2}^{\dag})],\label{eqc}\\
&\mathscr{L}_{2c}\hat{\rho}=&\sum^{2}_{i,j=1}\frac{\Gamma_{ijc}}{2}[(n_{2c}+1)(\hat{\sigma}_{gj}\hat{\rho}\hat{\sigma}_{gi}^{\dag}
                            +\hat{\sigma}_{gi}\hat{\rho}\hat{\sigma}_{gj}^{\dag}-\hat{\sigma}_{gj}^{\dag}\hat{\sigma}_{gi}\hat{\rho}-\hat{\rho}\hat{\sigma}_{gi}^{\dag}\hat{\sigma}_{gj})\nonumber\\
                            &&+n_{2c}(\hat{\sigma}_{gi}^{\dag}\hat{\rho}\hat{\sigma}_{gj}+\hat{\sigma}_{gj}^{\dag}\hat{\rho}\hat{\sigma}_{gi}
                                 -\hat{\sigma}_{gi}\hat{\sigma}_{gj}^{\dag}\hat{\rho}-\hat{\rho}\hat{\sigma}_{gj}\hat{\sigma}_{gi}^{\dag})],\label{eqd}\\
&\mathscr{L}_{3c}\hat{\rho}=&\frac{\gamma_{e}}{2}[(n_{e}+1)(2\hat{\sigma}_{21}\hat{\rho}\hat{\sigma}_{21}^{\dag}-\hat{\sigma}_{21}^{\dag}\hat{\sigma}_{21}
                              \hat{\rho}-\hat{\rho}\hat{\sigma}_{21}^{\dag}\hat{\sigma}_{21})\nonumber\\
                            &&+n_{e}(2\hat{\sigma}_{21}^{\dag}\hat{\rho}\hat{\sigma}_{21}-\hat{\sigma}_{21}\hat{\sigma}_{21}^{\dag}\hat{\rho}-\hat{\rho}\hat{\sigma}_{21}\hat{\sigma}_{21}^{\dag})],\label{eqe}\\
&\mathscr{L}_{\Gamma}\hat{\rho}=&\sum^{2}_{i=1}\frac{\Gamma_{i}}{2}(2\hat{\sigma}_{\alpha ii}\hat{\rho}\hat{\sigma}_{\alpha ii}-\hat{\rho}\hat{\sigma}_{\alpha ii}-\hat{\sigma}_{\alpha ii}\hat{\rho}),\label{eqf}\\
&\mathscr{L}_{\chi\Gamma}\hat{\rho}=&\sum^{2}_{i=1}\frac{\chi\Gamma_{i}}{2}(2\hat{\sigma}_{bi}\hat{\rho}\hat{\sigma}_{bi}^{\dag}-\hat{\rho}\hat{\sigma}
                                _{\alpha ii}-\hat{\sigma}_{\alpha ii}\hat{\rho}). \label{eqg}
\end{eqnarray}
\end{widetext}}

Here Eq.~(\ref{eqb}) describes the effect of high-temperature reservoir with \(n_{h}\) denoting its average photon number. The low temperature reservoir has the average phonon number \(n_{1c}=[\exp(\frac{(E_{e_{2}}-E_{\alpha_{i}})}{k_{B}T_{a}})-1]^{-1}\) in Eq.~(\ref{eqc}), where \(\gamma_{iic}\)=\(\gamma_{ic}(\gamma_{jjc}\)=\(\gamma_{jc})\) are the spontaneous decay rates from level \(|e_{2}\rangle\) to level \(|\alpha_{i}\rangle(i=1,2)\), respectively, and the cross-coupling \(\gamma_{ijc}\) with \(\gamma_{ijc}\)=\(\gamma_{jic}\) describes the effect of Fano interference. Similarly, another low temperature reservoir is described by Eq.~(\ref{eqd}) with \(n_{2c}\)=\([\exp(\frac{(E_{\beta_{i}}-E_{g})}{k_{B}T_{a}})-1]^{-1}\) being the cold reservoir phonon number. Here \(\Gamma_{ijc}\)=\(\Gamma_{jic}\) is defined as \(\Gamma_{ijc}\!=\!\eta_{1}\sqrt{\Gamma_{ic}\Gamma_{jc}}\) with \(\eta_{1}\) denoting the quantum interference robustness to describe the Fano interference induced by the spontaneous decay rates, \(\Gamma_{iic}\)=\(\Gamma_{ic}(\Gamma_{jjc}\)=\(\Gamma_{jc})\)(\(i, j=1,2\)), from level \(|\beta_{i,(i\!=\!1,2)}\rangle\) to level \(|g\rangle\). In Eq.~(\ref{eqe}), \(n_{e}\)=\([\exp(\frac{(E_{e_{1}}-E_{e_{2}})}{k_{B}T_{a}})]^{-1}\) is the corresponding thermal occupation number of photons at temperature \(T_{a}\). \(\hat{\sigma}_{\alpha_{ii}}\) in Eqs.~(\ref{eqf}) and (\ref{eqg}) is defined as \(\hat{\sigma}_{\alpha_{ii}}\)=\(|\alpha_{i}\rangle\langle\alpha_{i}|_{(i=1,2)}\). Next, 1 million training data sets within 100 fs are collected from the density matrix element equations (see the Appendix) with parameters\cite{LiZhao2021} listed in Tab.~(\ref{Table1}).

\begin{table}
\begin{center}
\caption{{\scriptsize Parameters used in our numerical calculations.}}
\label{Table1}
\vskip 0.3cm\setlength{\tabcolsep}{0.2cm}
\begin{tabular}{ccc}
\hline
\hline
                                                                        & Values                 & Units  \\
\hline
\(E_{e_{1}}-E_{g}=E_{e_{2}}-E_{g}\)                                     & 0.185                  & eV  \\
\(E_{e_{1}}-E_{\alpha_{1}}=E_{e_{2}}-E_{\alpha_{2}}\)                   & 0.2                    & eV  \\
\(E_{\beta_{1}}-E_{g}=E_{\beta_{2}}-E_{g}\)                             & 0.2                    & eV  \\
\(\gamma_{h}\)                                                          & \(2.48\times10^{-6}\)  & eV  \\
\(\gamma_{e}\)                                                          & 0.025                  & eV  \\
\(\gamma_{1c}=\gamma_{2c}\)                                             & 0.012                  & eV  \\
\(\Gamma_{1}=\Gamma_{2}\)                                               & 0.124                  & eV  \\
\(\Gamma_{1c}=\Gamma_{2c}\)                                             & 0.0248                 & eV  \\
\(T_{a}\)                                                               & 0.026                  & eV  \\
\(n_{h}\)                                                               & 6000                   &     \\
\(n_{e}\)                                                               & 0.46                   &     \\
\(\chi\)                                                                & 0.2                    &     \\
\(\eta_{1}\)                                                            & 0.25                   &     \\
\hline
\hline
\end{tabular}
\end{center}
\end{table}

\subsection{Multi-optimized recurrent neural networks(MRNN)}

\begin{figure}[htp]
\centering
\includegraphics[width=0.70\columnwidth] {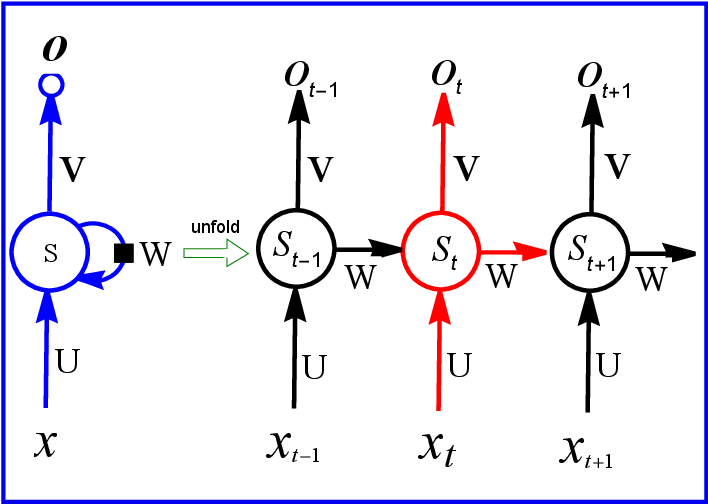}
\hspace{0in}%
\caption{{\scriptsize The architecture of the simple recurrent neural network (RNN) model.}}
\label{fig2}
\end{figure}

In this work, the dynamic evolution of populations on each energy level in Fig.~\ref{fig1} was utilized to illustrate EET evolution, and the long-term one in the proposed PSII-RC will be predicted using a recurrent neural network (RNN) tuned by four hyperparameters. The simple RNN\cite{2018Modelling,qu2022_resonance} is a type of neural network that is designed to learn data sequences such as time series, as illustrated in Fig.~\ref{fig2}. To further understand how they function, it is reasonable to compare them to more traditional feedforward neural networks, in which the input data set is propagated step by step through multiple intermediary layers with the training performed by updating the weight matrices and the vectors consecutively until the final output layer is reached. In this way the neural network learns the possible input-output correlations hidden behind the data, which can theoretically be used to handle temporal data.

However, a feedforward neural network is hardly the best solution since the number of free parameters rapidly grows with the number of time steps. As an alternative option, RNNs handle this issue by adopting a cyclic connection design in which the update rule for the hidden layers at time \(t\) is decided not only by the current state \(S_{t}\), but also by the previous state \((t-1)\), as shown in Fig.~\ref{fig2}, where \(x_i (i=t-1, t, t+1)\), \(S_{i}\) and \(O_{t_i}\) represent the input sample data, the memory of the input sample, and the state information stored at the time \(t_i\), respectively. Such a RNN update rule is given by two activation functions \(f\) and \(g\) written as

\begin{align}
&S_{t}=f (U x_{t}+W S_{t-1}),\\
&O_{t}=g(V S_{t}),\label{SO}
\end{align}

\noindent where the input sample weight \(U\), the input weight \(W\), and the output sample weight \(V\) are all time-independent matrices as the data set propagates forward. Here the activation functions \(f\) can be tanh, relu, sigmoid or other functions, while \(g\) is usually softmax and other functions. As shown in Eq.~(\ref{SO}), the previous state $S_{t-1}$ will participate the prediction of the current state $S_{t}$. In practice, using the three weight matrices, the input data and the hidden state at the previous time step can be taken into consideration in generating the hidden state at the current time step\cite{1997Bidirectional,2018Reinforcement}.

Due to these features of repeating operations, the Markovian assumption of independent data points at multiple time steps will become problematic in the application of RNN. If a machine learning model favors noise and minutiae, and ignores the general trends and patterns in training the data set, it will perform well on training data but poorly on generating new data, a phenomenon termed overfitting. To avoid this, we use a simple RNN with a multi-optimizer to predict the quantum evolutions of EET in light-harvesting complexes. Besides, the multi-optimized hyperparameters are addressed further below.

\subsubsection{Learning rate regulator (LRR)}

In each iteration, usually a learning rate regulator (LRR) is fitted to the simple RNN in order to update the model parameters and determine how far down the gradient the parameter moves with each update\cite{Ning9794728,Zhang10049413}. The exponential attenuation regulator, cosine LRR, preheating regulator, and LR attenuation regulator are all commonly used LRRs. A successful LRR should be able to optimize the model while avoiding overfitting and underfitting. If the LRR is set too high or too low, the model may diverge or converge slowly, leading to more training rounds before the optimal solution is obtained.

To simplify the task, we define an LRR at the starting stage of training. During training, the model computes the gradient of the loss function so as to determine the updating direction of the parameters, and such a procedure is repeated and updated again and again during the dynamic process. The performance of the model is evaluated using both the training and verification sets. Because the time series for the quantum evolutions of EET are projected to be long, a faster decaying LRR, i.e., an exponentially decaying LRR,

\begin{align}
LRR = 0.001\times \exp(\frac{- epoch}{10}),
\end{align}

\noindent is employed to this RNN. Here $\emph{epoch}$ is the parameter reflecting the iterations during the training.

\subsubsection{Early stop function}

In addition to the previously stated anti-overfitting strategies, an early stop function\cite{munoz2021rhoaso} is employed to decrease the occurrence of overfitting. As implied by its name, before the algorithm overfits, the early stop technique completes the training and obtains the optimal global outcome, resulting in a robust generalization performance, as shown by

\begin{align}
E_{opt}(t):=min_{t'}<=E_{va}(t')\label{eq11},\\
GL(t)=100\times[\frac{E_{va}(t)}{E_{opt}(t)}-1],\label{eq12}
\end{align}

\noindent where $E_{opt}(t)$ is the ideal verification error set as a function of the number of repetitions $t$, and \(GL(t)\) is the generalization loss evaluating the rate at which the generalization error grows in comparison to the previous lowest error. When the generalization error is large, an early stop is preferable since it indicates that the model has been fitted. Such an ending of training is judged by a threshold of \(GL(t)\). The early stop technique's halting criterion is classified into three types: the first, second, and third. In this work, the first type of stop rule is employed to determine the loss function and accuracy on the verification set.

\subsubsection{Regularization}

Regularization via adding penalty terms into the loss function of the model is becoming a popular strategy for reducing the overfitting and improving the model generalization in machine learning. There are two types of regularization, namely L1 and L2\cite{krueger2016zoneout,merity2017revisiting}, evaluated by two weight parameters $w^{*}$ and \(w\) following and before the update, respectively. They are given by

\begin{align}
w^{*}_{L1}=argmin _{w} \{MSE(y,\hat{y},w)+\frac{\lambda}{2} \sum_{i=1}^{n} |w_{i}|\}, \label{eql1}\\
w^{*}_{L2}=argmin _{w} \{MSE(y,\hat{y},w)+\frac{\lambda}{2} \sum_{i=1}^{n} {w_{i}}^2\},\label{eql2}
\end{align}
\begin{figure}[htp]
\centering
\includegraphics[width=0.75\columnwidth] {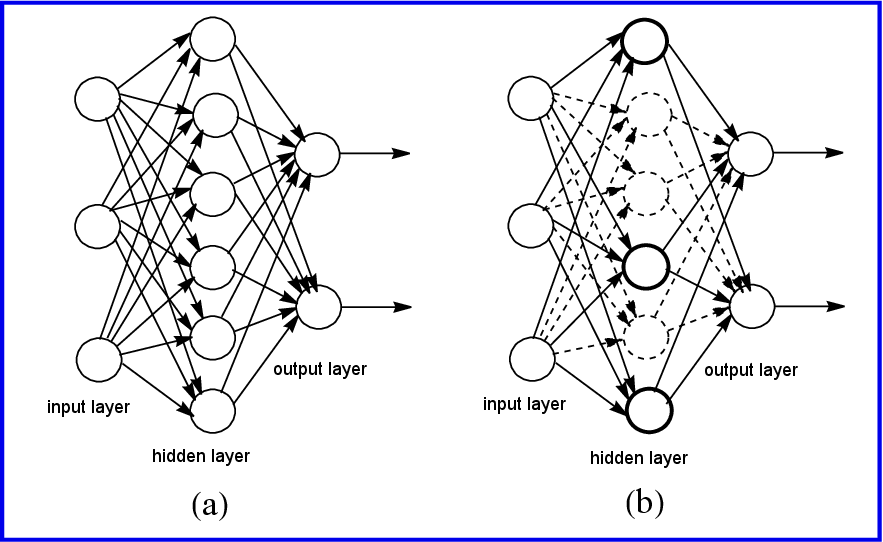}
\hspace{0in}%
\caption{{\scriptsize The structural diagram of Recurrent Neural Network (a) without and (b) with Dropout regularization techniques.}}
\label{fig3}
\end{figure}

\noindent where the sum of the absolute values and the square of the parameters have been added to the loss function, respectively\cite{merity2017revisiting}. The first terms on the right side of the two equations represent the original loss functions, the second terms represent the regularization parameters $\lambda$, and the summation is carried out over the numerous model ownership parameters. L1 regularization in Eq.~(\ref{eql1}) creates sparse solutions and is more suitable for feature selection, whereas L2 regularization Eq.~(\ref{eql2}) tends to produce a number of tiny but non-zero weights. In this study, L2 regularization is used to prevent overfitting in the long-term training data set.

\begin{table}
\begin{center}
\caption{Hyperparameters used in the RNNs training}
\label{Table2}
\vskip 0.2cm\setlength{\tabcolsep}{0.1cm}
\begin{tabular}{ccccc}
\hline
\hline
                                    &\(L2\)    &\(LR\)     &\(epochs\)   & Dropout     \\
\hline
\(\rho_{\alpha_{1}\alpha_{1}}\)     & 0.01     & 0.0001    & 139        & 0.11747474747474748 \\
\(\rho_{\alpha_{2}\alpha_{2}}\)     & 0.01     & 0.0001    & 140        & 0.09779519760654846  \\
\(\rho_{e_{1}e_{1}}\)               & 0.01     & 0.0000001 & 155        &0.32499999999999996 \\
\(\rho_{e_{2}e_{2}}\)               & 0.01     & 0.0000001 & 125        &0.32499999999999996  \\
\(\rho_{bb}\)                       & 0.01     & 0.0000001 & 122        &0.35                 \\
\hline
\hline
\end{tabular}
\end{center}
\end{table}

\subsubsection{Dropout}

In this work, the Dropout regularization technique\cite{gajbhiye2018exploration,sarma2021structured,salehin2023review} is also applied to the simple RNN. As the training of the input data is carried out forward through the neural network, the estimated loss propagates backward, and the parameters are changed using the gradient descent approach, as shown in Fig.~\ref{fig3}(a). In Fig.~\ref{fig3}(b), the Dropout parameter \(P\) is introduced to deactivate certain neurons with a specific probability during the forward propagation, and the parameters are updated using the gradient descent approach. This method is repeated again and again so as to properly alleviate overfitting during the training process in this proposed MRNN.

\section{Results and discussions}
\subsection{Training the multi-optimized recurrent neural network (MRNN)}

\begin{figure}[htp]
\centering
\includegraphics[width=0.85\columnwidth]{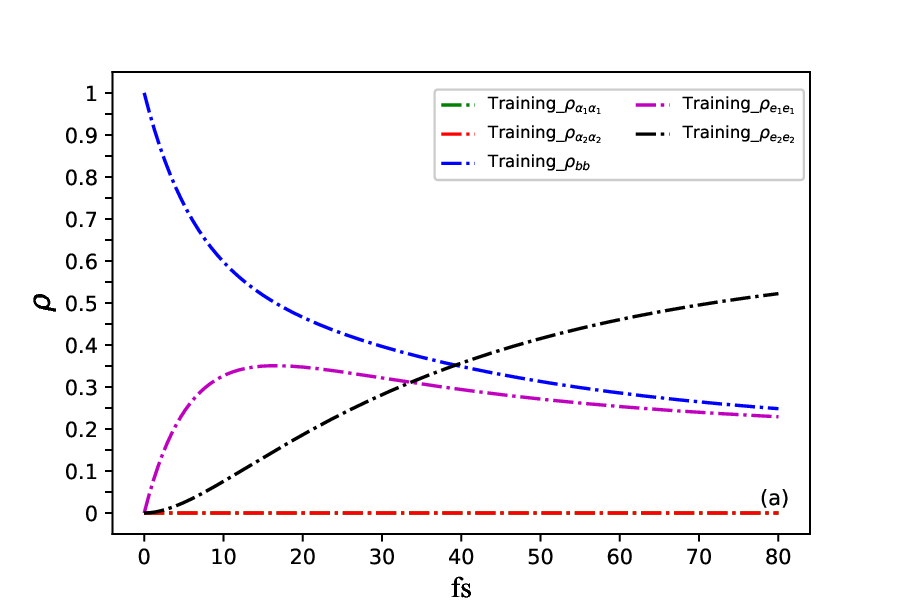}\hspace{0in}%
\includegraphics[width=0.85\columnwidth] {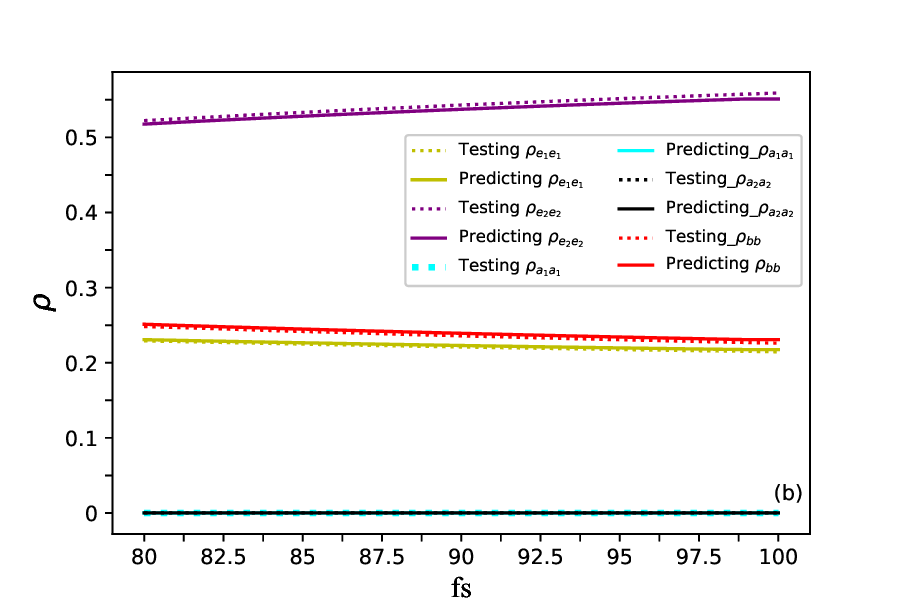}
\hspace{0in}%
\caption{{\scriptsize (Color online) (a) Evolutions of the training population at all excited states within the interval[0,~80]fs. (b) A comparison of learning quality between our proposed learning model and the collected testing set over the range [80,100]fs. The hyperparameters listed in Tab.(\ref{Table2}) and the original codes in \href{https://pan.baidu.com/s/1w9fXbLA0i_IgOV2cvsTzQg?pwd=f3n8}{SM1, SM2 and SM3} are used for the calculation.}} \label{fig4}
\end{figure}

 After progressively incorporating a few hyperparameters, such as early stop function, L2 regularization method, LRR, Dropout, and Bayesian optimizer, a multi-optimized RNN model is constructed. As a distinguishing feature, this model utilizes the Lindblad-type master equation Eq.~(\ref{eqa}) to gather data from the proposed PSII-RC, with parameters listed in Tab.~(\ref{Table1}). By using Eq.~(\ref{eqa}), we produce a data set of 1 million data points in 100 fs, which are divided into training and test sets at 4:1 ratio. The first 800,000 data points serve as the training set, while the remaining 200 000 act as the test set, using the hyperparameters listed in Tab.~(\ref{Table2}). Fig.~\ref{fig4}(a) displays the evolution of the training set fed into this proposed multi-optimized RNN learning model (\href{https://pan.baidu.com/s/1w9fXbLA0i_IgOV2cvsTzQg?pwd=f3n8}{Original codes in SM1}) during the interval [0,~80]fs.

In order to assess the validity of the proposed learning model, we predict the evolution of EET (solid curves) over a time range from 80 to 100 fs with a comparison to the test set (dotted curves) collected from the aforementioned PSII-RC within the same range, indicating a good agreement between them, as illustrated in Fig.~\ref{fig4}(b). This ensures the high accuracy of this learning model composed of the simple RNN and some optimizers(\href{https://pan.baidu.com/s/1w9fXbLA0i_IgOV2cvsTzQg?pwd=f3n8} {Original codes in SM1}). Supported by these results, the proposed multi-optimized RNN learning model is expected to be able to anticipate the evolution of EET from 80 to 500 fs.

\subsection{EET Predicted by the MRNN}

\begin{figure}[htp]
\centering
\includegraphics[width=0.59\columnwidth] {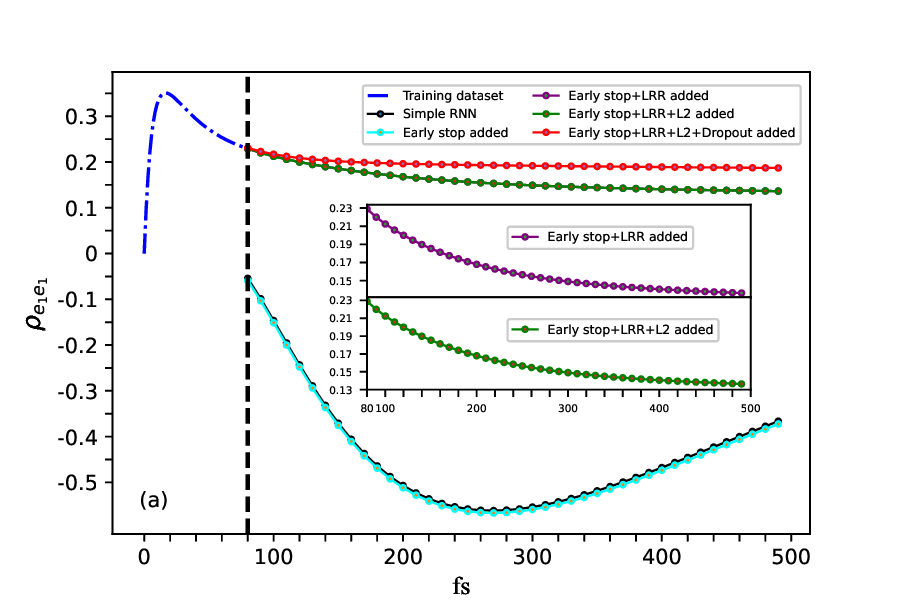}\hspace{0in}%
\includegraphics[width=0.59\columnwidth] {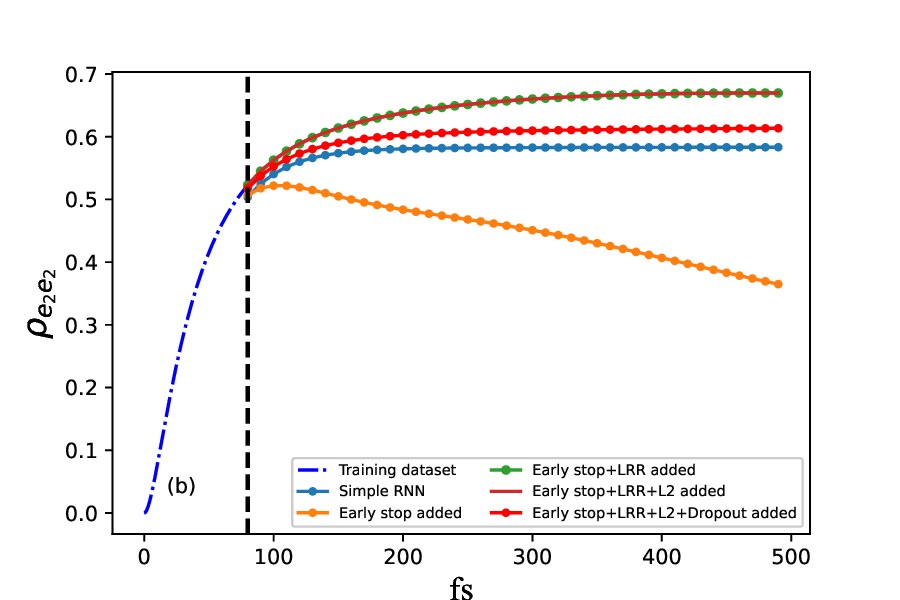}\hspace{0in}%
\includegraphics[width=0.59\columnwidth] {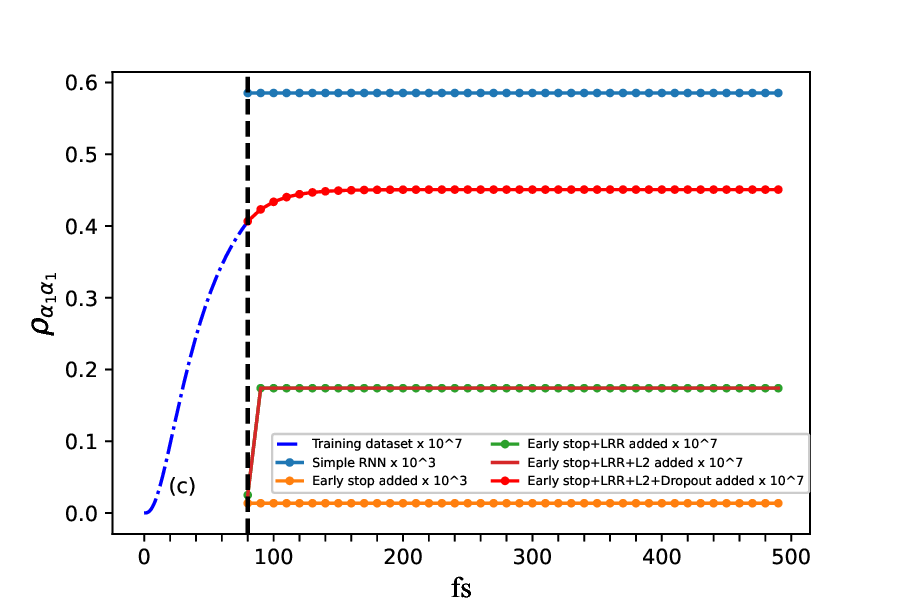}\hspace{0in}%
\includegraphics[width=0.59\columnwidth] {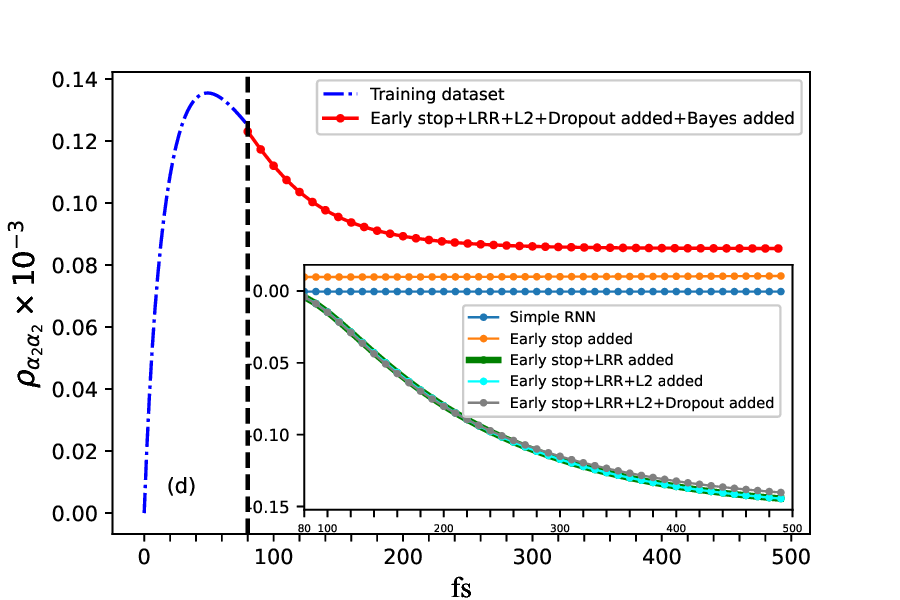}\hspace{0in}%
\includegraphics[width=0.59\columnwidth] {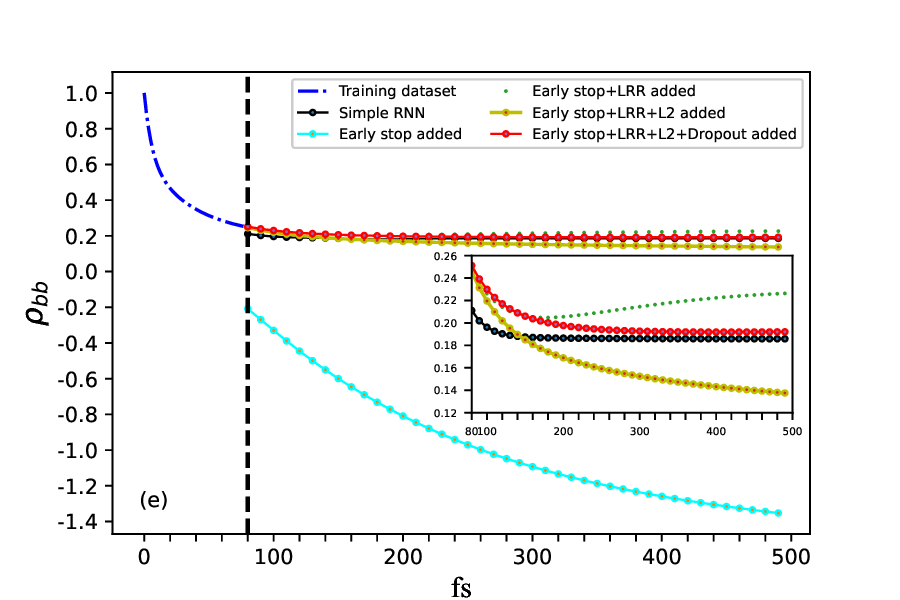}\hspace{0in}%
\includegraphics[width=0.59\columnwidth] {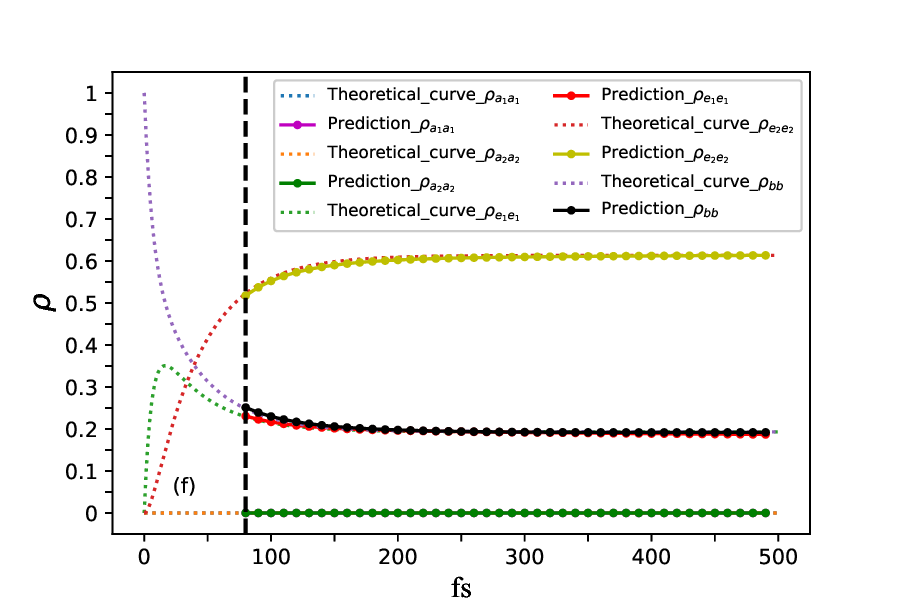}\hspace{0in}%
\caption{{ \scriptsize (Color online) Prediction of the evolutive EET with different combinations of hyperparameters are shown in (a) to (e) via the multi-optimized RNN in the interval [80, 500]fs, and the predictive and theoretical (within [0, 500] fs) EET evolutions are demonstrated in (f) with all identical parameters to those in Fig.\ref{fig4} included in the calculation.}} \label{fig5}
\end{figure}

In Fig.~\ref{fig5}, the vertical dashed black lines at the 80th fs shows the temporal starting point of our prediction achieved by gradually incorporating hyperparameters into the calculation. The cut-off at 80 fs is used to evaluate the prediction accuracy by checking how precisely the data generated by the prediction coincide with the training data set. After examining Figs.~\ref{fig5}(a)-(e), the evolutionary history of each population indicates that our predictions vary against hyperparameters in the following periods. L2 regularization, on the other hand, has little effect on EET prediction, as shown by the nearly identical curves in Figs.~\ref{fig5}(a)-(e). When all the hyperparameters are added into the simple RNN, there is a notable consistence between the training and prediction values at 80 fs, corresponding to the red curves in Figs.~\ref{fig5}(a)-(e)(\href{https://pan.baidu.com/s/1w9fXbLA0i_IgOV2cvsTzQg?pwd=f3n8}{Original codes in SM2}).

Consider the prediction of EET on the excited state \(|e_{1}\rangle\) in Fig.~\ref{fig5}(a), where the hyperparameters and optimizers are added one by one within the RNN architecture. Three layers with 128, 64, and 32 neurons, respectively, are the essential characteristic neural network architecture. In the absence of optimizers and hyperparameters, the simple RNN timing prediction model fails to generate substantial physical results, as the prediction (black dotted solid line) is not consistent with the training curve at 80 fs in Fig.~\ref{fig5}(a). The time forecast equipped by the subsequently added early stop function does not make any difference to this inconsistence, as seen by the sky blue dotted solid curve overlapping that for simple RNN. The overlapping purple and green curves in Fig.~\ref{fig5}(a) indicate that the evolution of EET is not sensitive to the L2 regularization on the excited state \(|e_{1}\rangle\), and almost the same evolutionary properties over [80,~500]fs were exhibited by the insets in Fig.~\ref{fig5}(a). The evolution curve of \(|e_{1}\rangle\) tends to level off when the Dropout parameter \(P\)=0.32499999999999996 is applied for further optimization, as demonstrated by the red curve.

In Fig.~\ref{fig5}(c), the red curve achieves perfect docking with the training data at 80 fs with the Dropout parameter set as \(P\)=0.11747474747474748 and the initial LR= 0.001. As it takes time to manually adjust the hyperparameters of the neural network when optimizing the prediction of the dynamical population on \(|\alpha_{2}\rangle\), the Bayesian optimizer is implemented into the neural network to find the ideal number of layers and neurons. Finally, a neural network composed of 6.23223034545932 layers with 124.19253861421566 neurons per layer was employed, corresponding to the red curve shown in Fig.~\ref{fig5}(d). The same neural network architecture parameters were employed to forecast \(|b\rangle\), and the inset in Fig.~\ref{fig5}(e) clearly shows the roles each hyperparameter plays in predicting the evolution of \(\rho_{bb}\).

\begin{figure}[htp]
\centering
\includegraphics[width=0.8\columnwidth] {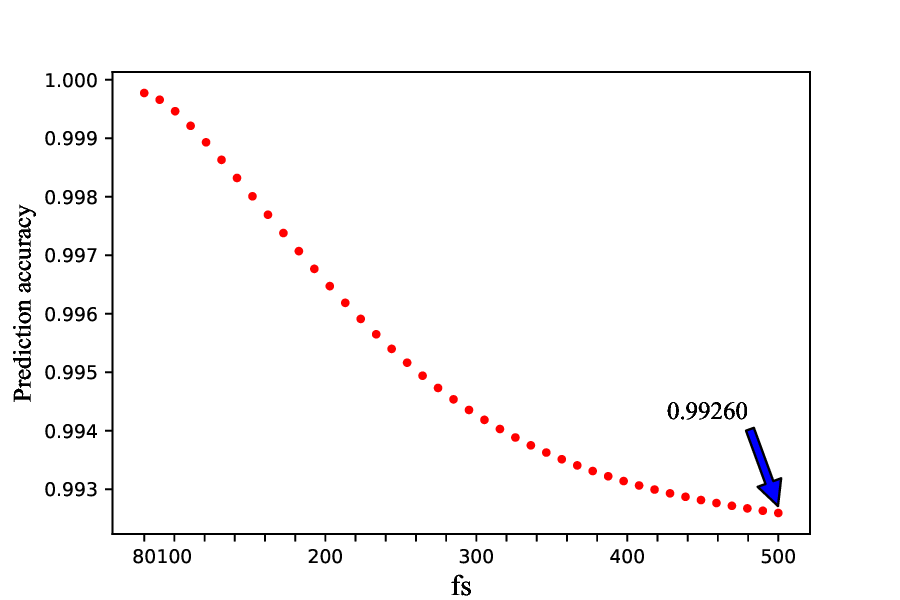}\hspace{0in}%
\caption{{\scriptsize (Color online) Prediction accuracy defined by the normalized population sum in the interval [80,~500]fs.}}\label{fig6}
\end{figure}

To validate the predictive accuracy of the proposed MRNN in this work, the evolution behaviors of EET are theoretically simulated by using Eq.~(\ref{eqa}) in the temporal range of [0,~500] fs, as shown by the dotted lines in Fig.~\ref{fig5}(f), while the optimal prediction results are re-plotted as dotted solid curves in the time range 80 to 500 fs, as shown by the red curves in Figs.~\ref{fig5}(a)-(e) for comparison(\href{https://pan.baidu.com/s/1w9fXbLA0i_IgOV2cvsTzQg?pwd=f3n8}{Original codes in SM1}). It is found that the theoretical calculation and the curves predicted by MRNN for photosynthetic EET perfectly coincide, indicating the validity of our proposed MRNN.

In order to quantitatively evaluate the prediction precision of EET for the system, an accuracy rate is defined as the normalized population sum during [80,500 fs]. Although the prediction precision decreases against time for this photosynthetic system, it reaches a low of 0.9926 at 500fs, as shown by an arrow in Fig.~\ref{fig6}(\href{https://pan.baidu.com/s/1w9fXbLA0i_IgOV2cvsTzQg?pwd=f3n8}{Original codes in SM1}). Nevertheless, this accuracy outperforms most RNN learning models, such as those reported by Refs.~\cite{2021Modelling,LI2023148}. At the same time, because our MRNN learning model is born from the simple RNN, long-term prediction inevitably will reveal the inherent flaw of short-term prediction, resulting in a progressive drop in accuracy, as seen by the decreasing feature of Fig.~\ref{fig6}.

\subsection{ Polynomial fitting and analytical expression for the predictive EET}

\begin{figure}[htp]
\centering
\includegraphics[width=0.7\columnwidth] {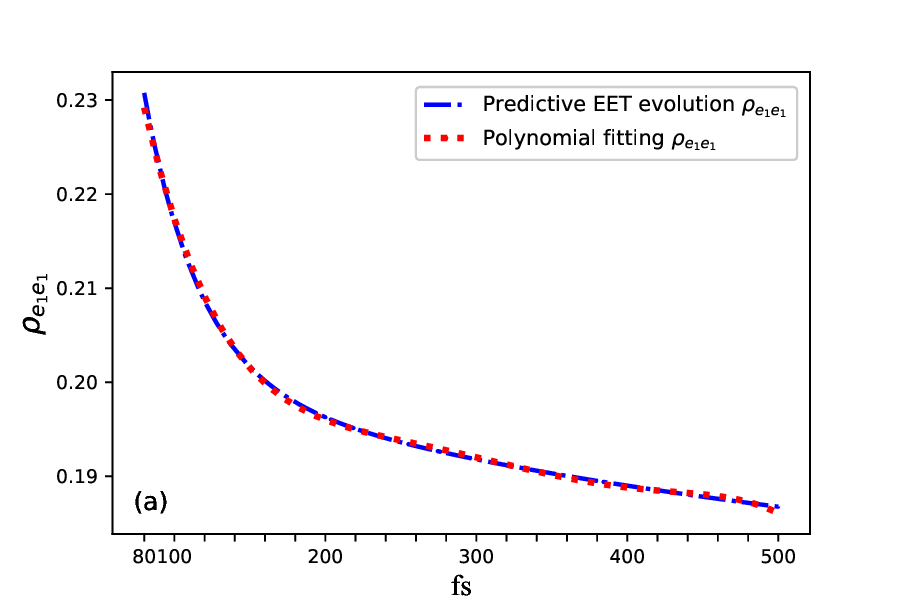}\hspace{0in}%
\includegraphics[width=0.7\columnwidth] {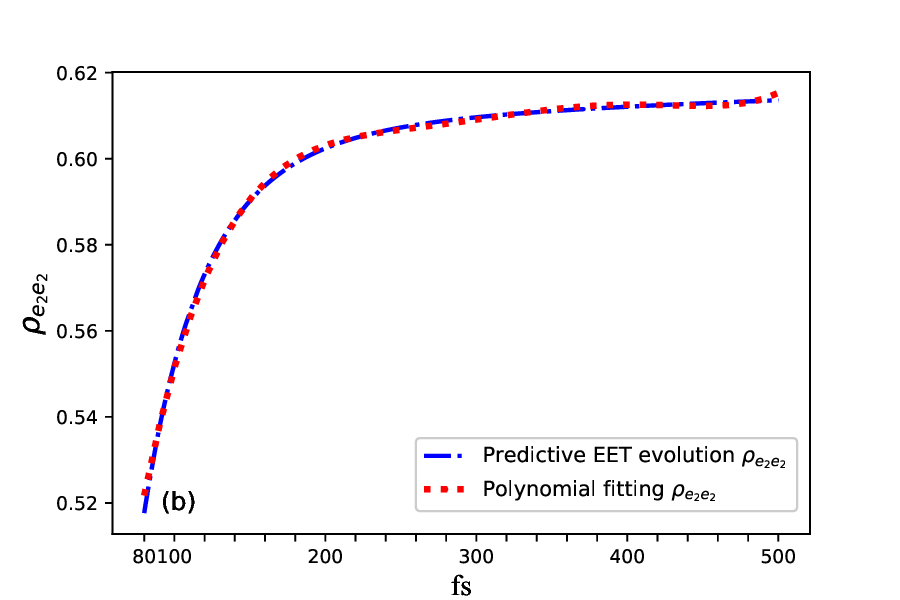}\hspace{0in}%
\includegraphics[width=0.7\columnwidth] {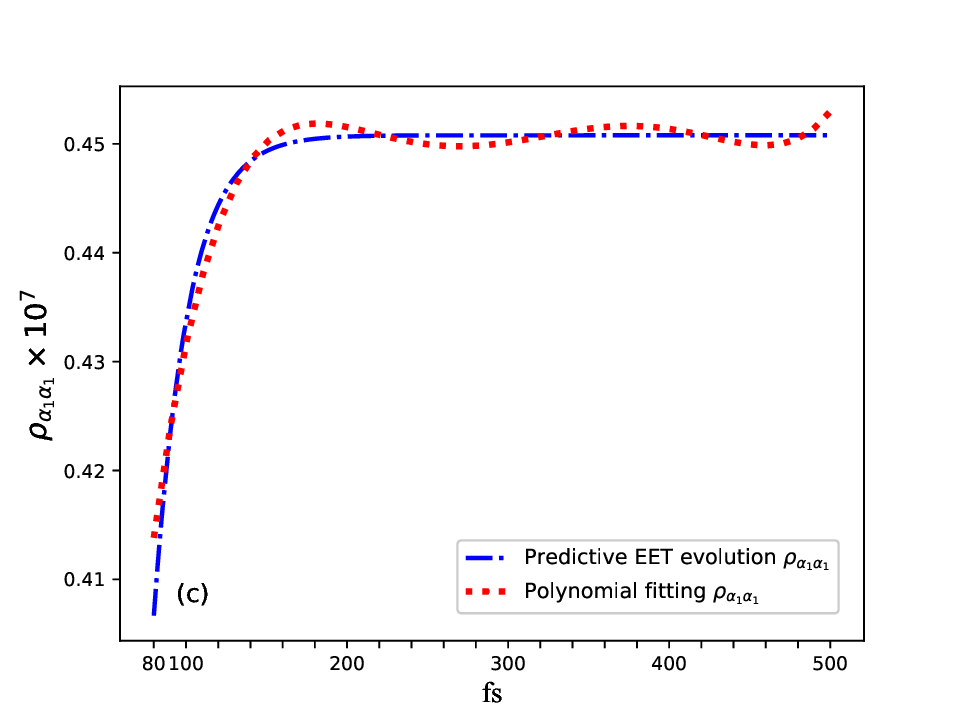}\hspace{0in}%
\includegraphics[width=0.7\columnwidth] {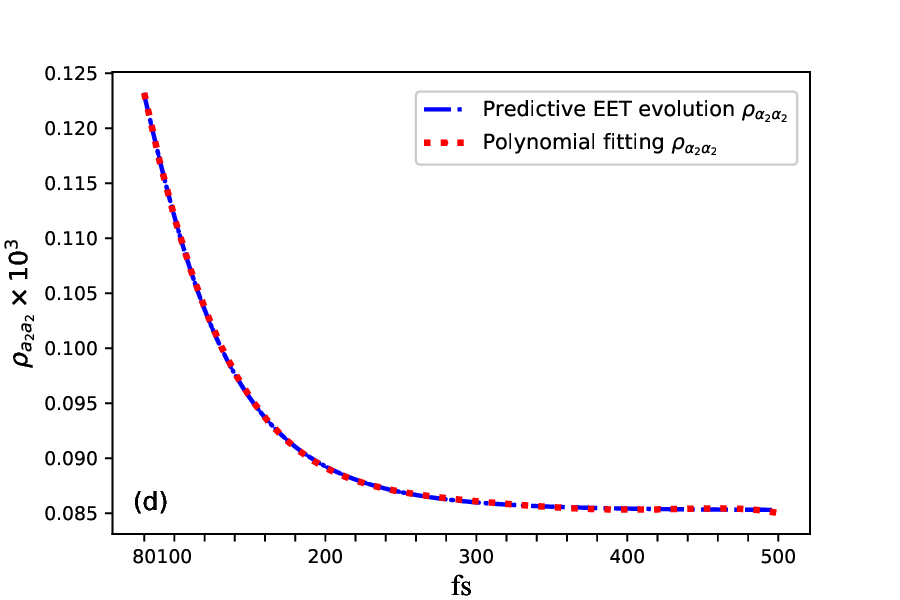}\hspace{0in}%
\includegraphics[width=0.7\columnwidth] {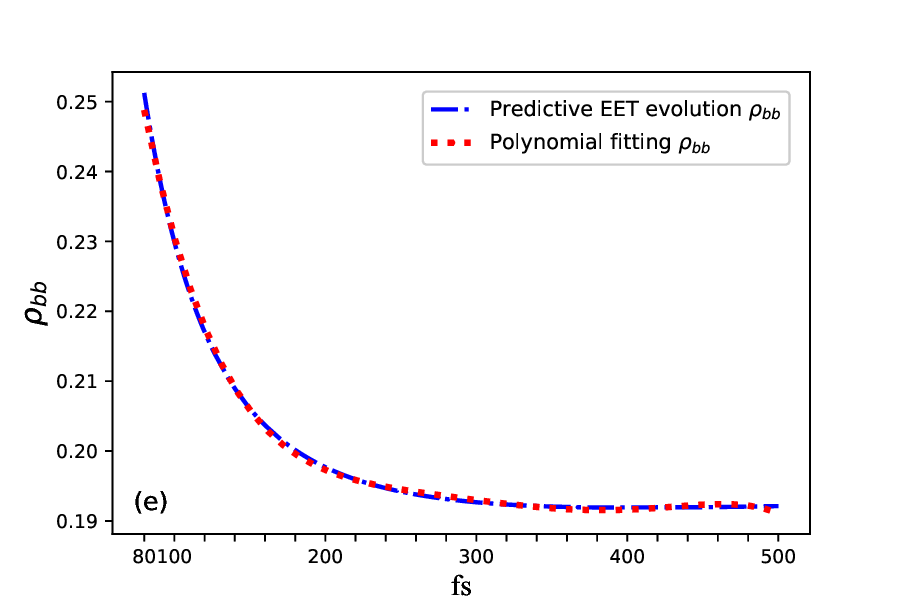}
\hspace{0in}%
\caption{{\scriptsize (Color online) Polynomial fittings (red dotted curves) in comparison with EET evolution predictions (blue dash-dotted curves) within the interval of [80, 500]fs.}}
\label{fig7}
\end{figure}

Using a polynomial to fit the predicted findings is a useful mathematical method for gaining an analytical knowledge of EET. Such a fitting usually relies on the least squares method and lowering the error sum of squares\cite{pei2005mapping,escobedo2016neural}. In Python, the Polynomial Features function is a utility in the scikit-learn library to generate the polynomial features so as to form a new feature matrix. Finally, the feature matrix is returned for further model training and fitting. When a polynomial of order 2 is provided for a one-dimensional feature $A$, Polynomial Features generate a new feature matrix containing $A$ to the first and second powers. If the original feature has many dimensions, the resulting feature matrix will have power combinations for each of them.

Figure~\ref{fig7} displays the differences between the predicted curves and the fitting curves using Polynomial Features in scikit-learn library. A maximum degree of order 5 for all polynomial fittings is adopted in Fig.~\ref{fig7} (\href{https://pan.baidu.com/s/1w9fXbLA0i_IgOV2cvsTzQg?pwd=f3n8}{Original codes in SM3}).  As a result, the analytical polynomial fittings corresponding to to Figs.~\ref{fig7}(a)-(e), respectively, are found to be

{\footnotesize
\begin{widetext}
\begin{eqnarray}
&\rho_{e_{1}e_{1}}(t) \!=\! &-0.0022184223655202433 t +1.3936336816327624 \times10^{-5}t^{2}         \nonumber\\
                        &&-4.3658376701560287 \times10^{-8}t^{3}+6.694506984131638 \times10^{-11}t^{4}\nonumber\\
                        &&-4.008815848721703 \times10^{-14}t^{5}+0.33721254828307511,  \label{16}\\
&\rho_{e_{2}e_{2}}(t) \!=\! & 0.0055891350945578 t-3.50638450439065\times10^{-5} t^{2}\nonumber\\
                       &&+1.091124267760133 \times10^{-7} t^{3}-1.6659867620310589 \times10^{-10} t^{4}       \nonumber\\
                       &&+9.953236151938327 \times10^{-14} t^{5}+0.24964385363485108,                                   \label{17}\\
&\rho_{\alpha_{1}\alpha_{1}}(t)\!=\! &4.317228690621229\times10^{-10} t -3.0372753727222955 \times10^{-12} t^{2}\nonumber\\
                                      &&+1.0210132754732106 \times10^{-14} t^{3}-1.647224363731816 \times10^{-17} t^{4}\nonumber\\
                                      &&+1.0249801765297704 \times10^{-20} t^{5}+0.00000002169596502,	               \label{18}\\
&\rho_{\alpha_{2}\alpha_{2}}(t)\!=\! &-1.8447722039325343 \times10^{-6} t+1.0562606143359824 \times10^{-8} t^{2}     \nonumber\\
                                &&-3.040707800147938 \times10^{-11}t^{3}+4.356557804662399 \times10^{-14} t^{4}     \nonumber\\
                                &&-2.4712186236144212 \times10^{-17} t^{5}+0.00021705673079966,                        \label{19}\\
&\rho_{bb}(t)\!=\! &-0.003319998227995093 t+2.0476938571365984\times10^{-5}  t^{2}                                 \nonumber\\
                  &&-6.313410066086038 \times10^{-8} t^{3}+9.606627816067419 \times10^{-11} t^{4}          \nonumber\\
                  &&-5.735689700969715 \times10^{-14} t^{5}+0.41192131944573884,                                        \label{20}
\end{eqnarray}
\end{widetext}}

\noindent where \(\rho_{e_{1}e_{1}}\), \(\rho_{e_{2}e_{2}}\), \(\rho_{\alpha_{2}\alpha_{2}}\), and \(\rho_{bb}\) agree very well with the predicted curves, as shown in Fig.~\ref{fig7}. In contrast, the fitting result of \(\rho_{\alpha_{1}\alpha_{1}}\) is rather poor, as shown in Fig.~\ref{fig7}(c), where overfitting is visible in its dynamic population progress, indicating that an alternative fitting technique should be found for it.

Because the characteristic features of an RNN lies in its internal (hidden) loop memory, it is understandable that a dynamic state contains all the information about the previous input as it evolves through the data sequence, as shown in Fig.~\ref{fig2}. If the polynomial order is increased, the polynomial features will over adapt to the past information but perform poorly on the fresh data. This enlarges both amplitude and rate of the change, leading to overfitting. An effective solution is to derive information based on not just purely the previous step, but all the previous inputs, by implementing a time-segment strategy. Here polynomial fittings in different time ranges are carried out and a unit step function

{\footnotesize
\begin{eqnarray}
f_{1}(t)=\left
\{
\begin{array}{cl}
1 & \quad t<250\\
0 & \quad t>=250\\
\end{array} \right.
,f_{2}(t)=\left
\{
\begin{array}{cl}
0 & \quad t<250\\
1 & \quad t>=250\\
\end{array} \right.\label{21}
\end{eqnarray}}

\noindent is incorporated to the final polynomial expression in each time-segment. The fitting time is divided into two intervals [80,~250] fs and [250,~500] fs, and the polynomial fitting in each interval is multiplied by a unit step function given by Eq.~(\ref{21}). Their final fitting is the sum of the two polynomial fittings, i.e.

{\footnotesize
\begin{eqnarray}
&\rho_{\alpha_{1}\alpha_{1}}(t_{80-500})=&\rho_{\alpha_{1}\alpha_{1}}(t_{80-250})*f_{1}(t)   \nonumber\\
                             &&+\rho_{\alpha_{1}\alpha_{1}}(t_{250-500})*f_{2}(t).            \label{22}
\end{eqnarray}}

\begin{figure}
\centering
\includegraphics[width=0.85\columnwidth] {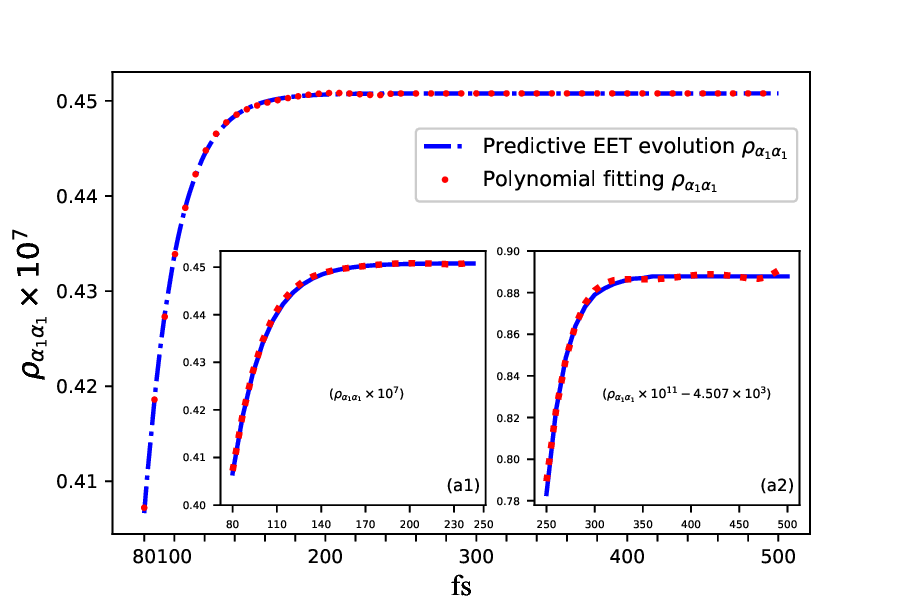}\hspace{0in}%
\caption{{\scriptsize (Color online) A comparison between the fittings by the time-segment polynomial multiplied by unit step function (red dotted curves) and the evolution predictions (blue dash-dotted curves) in [80,~500]fs. Time-segment polynomial fittings in comparison with the evolution predictions shown by the insets(a1) in [80,~250]fs and (a2) in [250,~500]fs.}}\label{fig8}
\end{figure}

The curves in Fig.~\ref{fig8} shows the fitting results for the population on the state \(|\alpha_{1}\rangle\) using the aforementioned procedure (\href{https://pan.baidu.com/s/1w9fXbLA0i_IgOV2cvsTzQg?pwd=f3n8}{Original codes in SM3}). The two insets in Fig.~\ref{fig8} show a comparison between polynomial fitting and predictions in the intervals [80,250]fs and [250,500] fs. respectively. The overlapping between the dotted lines (red) and dash-dotted lines (blue) demonstrates the high precision of polynomial fitting at various intervals. Furthermore, the analytical functions derived by the time-segment polynomial fittings are, respectively, given by
{\footnotesize
\begin{eqnarray}
&\rho_{\alpha_{1}\alpha_{1}}(t_{80-250}) &=1.7388382452293818 *10^{-9}t              \nonumber\\
                                      &&-1.8538908793974676 *10^{-11}t^{2}           \nonumber\\
                                      &&+9.883817236997048 *10^{-14}t^{3}             \nonumber\\
                                      &&-2.6256355830257586 *10^{-16}t^{4}             \nonumber\\
                                      && +2.7737661365102853 *10^{-19}t^{5},           \label{23}\\
&\rho_{\alpha_{1}\alpha_{1}}(t_{250-500}) & =2.500773638209706*10^{-12}t           \nonumber\\
                                     &&-1.2744394675654058 *10^{-14}t^{2}          \nonumber\\
                                     && +3.227678944102217 *10^{-17}t^{3}           \nonumber\\
                                     &&-4.0616638358969016 *10^{-20}t^{4}            \nonumber\\
                                     &&+2.031479651834943 *10^{-23}t^{5}.              \label{24}
\end{eqnarray}}

The comparison made between the whole period fitting and evolution prediction in [80,~500]fs for the dynamics populations on state \(|\alpha_{1}\rangle\) demonstrates that the time-segment polynomial fitting can well overcome the overfitting in Fig.~\ref{fig7}(c).

\begin{figure}
\centering
\includegraphics[width=0.85\columnwidth] {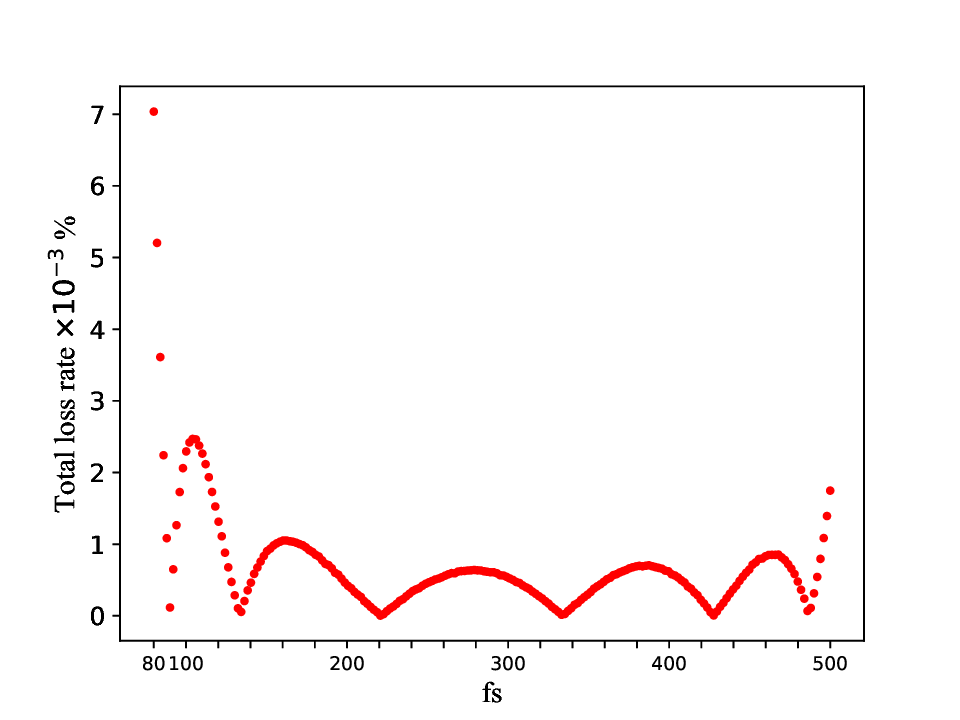}\hspace{0in}%
\caption{{\scriptsize (Color online) Total loss rate of the polynomial fitting versus the prediction within the range of [80,~500]fs.}}\label{fig9}
\end{figure}

Figure~(\ref{fig9}) depicts the total fitting loss rate compared to the predicted results, a physical quantity assessing the precision of the fitting technique employed in this work. It is found that even though a jumpy loss rate can be seen both at the initial and the final stages of the time interval, the loss rate exhibits an oscillating behavior of an order of $10^{-5}$, ensuring a high accuracy for this MRNN when utilized in this work.

So far, the above fitting findings have accurately described the EET evolution, revealing the physical rules behind data, which may be utilized to purposely control EET and construct artificial photosynthetic devices in the future.

\section{Conclusion and outlook}

In summary, fed by the original PSII-RC data set, a MRNN strategy is proposed to forecast the EET evolution, with an accuracy of over 99.26\% within 500fs if compared to the theoretical deduction. The polynomial fitting is also implemented for the EET evolutions so as to get analytical results. The predicted EET evolutions are also subjected to time-segment polynomial fitting multiplied by a unit step function, and the analytical formulations showing a high precision with the loss rate of an order of $10^{-5}$ demonstrate the closeness to physical law. The results reveal that the proposed MRNN is a valid and powerful data mining tool for forecasting the evolution of EET in a light-harvesting complex. A comparison to experimental data in the future is expected so as to assess the validity of this learning model.

is obtained.
\section*{Author contributions}

S. C. Zhao conceived the idea. Y. M. Huang performed the numerical computations and wrote the draft, and S. C. Zhao did the analysis and revised the paper. Z. R. Zhao gave some discussion in the numerical computations.

\section{Acknowledgments}

S. C. Zhao is grateful for the significant discussions and language polishing from Prof. Chen-Xu Wu of Xiamen Univ., and for fundings from the National Natural Science Foundation of China (grants 62065009 and 61565008) and Foundation for Personnel training projects of Yunnan Province, China (grant 2016FB009).

\section{APPENDIX}

The density matrix dynamic element equations are given by

\begin{widetext}
\begin{eqnarray}
&\dot{\rho}_{e_{1}e_{1}}=&-\gamma_{e}[(n_{e}+1)\rho_{e_{1}e_{1}}-n_{e}\rho_{e_{2}e_{2}}]-\gamma_{h}[(n_{h}+1)\rho_{e_{1}e_{1}}-n_{h}\rho_{gg}],\nonumber\\
&\dot{\rho}_{e_{2}e_{2}}=&\gamma_{e}[(n_{e}+1)\rho_{e_{1}e_{1}}-n_{e}\rho_{e_{2}e_{2}}]-\gamma_{1c}[(n_{1c}+1)\rho_{e_{2}e_{2}}-n_{c1}\rho_{\alpha_{1}\alpha_{1}}]\nonumber\\
                         &&-\gamma_{2c}[(n_{1c}+1)\rho_{e_{2}e_{2}}-n_{1c}\rho_{\alpha_{2}\alpha_{2}}]+2\gamma_{12c}n_{c1}Re[\rho_{\alpha_{1}\alpha_{2}}],\nonumber\\
&\dot{\rho}_{\alpha_{1}\alpha_{1}}=&\gamma_{1c}[(n_{1c}+1)\rho_{e_{2}e_{2}}-n_{1c}\rho_{\alpha_{1}\alpha_{1}}]-\gamma_{12c}n_{1c}Re[\rho_{\alpha_{1}\alpha_{2}}]-(1+\lambda)\Gamma_{1}\rho_{\alpha_{1}\alpha_{1}},\nonumber\\
&\dot{\rho}_{\alpha_{2}\alpha_{2}}=&\gamma_{1c}[(n_{1c}+1)\rho_{e_{2}e_{2}}-n_{1c}\rho_{\alpha_{1}\alpha_{1}}]-\gamma_{12c}n_{1c}Re[\rho_{\alpha_{1}\alpha_{2}}]-(1+\lambda)\Gamma_{2}\rho_{\alpha_{2}\alpha_{2}},\nonumber\\
&\dot{\rho}_{\alpha_{1}\alpha_{2}}=&-i\triangle_{1}\rho_{\alpha_{1}\alpha_{2}}-\frac{1}{2}(\gamma_{1c}+\gamma_{2c})n_{1c}\rho_{\alpha_{1}\alpha_{2}}+\frac{1}{2}\gamma_{12c}[2(n_{1c}+1)\rho_{e_{2}e_{2}}-n_{1c}\rho_{\alpha_{2}\alpha_{2}}-n_{1c}\rho_{\alpha_{1}\alpha_{1}}]\nonumber\\
&\dot{\rho}_{\beta_{1}\beta_{1}}=&\Gamma_{1}\rho_{\alpha_{1}\alpha_{1}}-\Gamma_{1c}[(n_{2c}+1)\rho_{\beta_{1}\beta_{1}}-n_{2c}\rho_{gg}]
                                 -\Gamma_{12c}(n_{2c}+1)Re[\rho_{\beta_{1}\beta_{2}}],\nonumber\\
&\dot{\rho}_{\beta_{2}\beta_{2}}=&\Gamma_{2}\rho_{\alpha_{2}\alpha_{2}}-\Gamma_{2c}[(n_{2c}+1)\rho_{\beta_{2}\beta_{2}}-n_{2c}\rho_{gg}]
                                 -\Gamma_{12c}(n_{2c}+1)Re[\rho_{\beta_{1}\beta_{2}}],\nonumber\\
&\dot{\rho}_{\beta_{1}\beta_{2}}=&-i\triangle_{2}\rho_{\beta_{1}\beta_{2}}-\frac{1}{2}(\Gamma_{1c}+\Gamma_{2c})(n_{2c}+1)\rho_{\beta_{1}\beta_{2}}-\frac{1}{2}\Gamma_{12c}[(n_{2c}+1)\rho_{\beta_{1}\beta_{1}}+(n_{2c}+1)\rho_{\beta_{2}\beta_{2}}
                                 -2n_{2c}\rho_{gg}],\nonumber\\
&\rho_{gg}=&1-\rho_{e_{1}e_{1}}-\rho_{e_{2}e_{2}}-\rho_{\alpha_{1}\alpha_{1}}-\rho_{\alpha_{2}\alpha_{2}}-\rho_{\beta_{1}\beta_{1}}-\rho_{\beta_{2}\beta_{2}},\nonumber
\end{eqnarray}
\end{widetext}

\noindent where \(\triangle_{1}=E_{\alpha_{1}}-E_{\alpha_{2}}\) and \(\triangle_{2}=E_{\beta_{1}}-E_{\beta_{2}}\) are the splitting of the states \(|\alpha_{1}\rangle(|\alpha_{2}\rangle)\) and \(|\beta_{1}\rangle(|\beta_{2}\rangle)\). We utilize the equations to simulate dynamics of EET in PSII-RC.

\section*{Data Availability Statement}

This manuscript has associated data in a data repository. [Authors' comment: All data included in this manuscript are available upon reasonable request by contacting with the corresponding author]. The Supporting Information is available free of charge at: \href{https://pan.baidu.com/s/1w9fXbLA0i_IgOV2cvsTzQg?pwd=f3n8} {Supplements to MRNN}.
\bibliography{reference}
\bibliographystyle{unsrt}

\end{document}